\documentclass[aps,prb,twocolumn]{revtex4}

\usepackage{color}
\usepackage{graphicx}
\usepackage{ulem}

\newcommand{\nrl}[1]{{#1}_{\mathcal{Z}}}

\newcommand{\ihq}{\frac{{{i}}}{{\hbar}}}
\newcommand{\EW}[1]{\left\langle #1 \right\rangle}
\newcommand{\sEW}[1]{{\mathbf{E}}\left[ #1 \right]}

\begin{document}

\title{Work and heat for two-level systems in dissipative environments:\\ Strong driving and non-Markovian dynamics}
\date{\today}
\author{R. Schmidt$^{1}$, M F. Carusela$^{2}$, J. P. Pekola$^{3}$,  S. Suomela$^{4}$ and J. Ankerhold$^{5}$}
\address{$^1$ School of Mathematical Sciences, University of Nottingham, University Park, NG7 2RD, United Kingdom\\
$^2$ Instituto de Ciencias, Universidad Nacional de General Sarmiento, J.M.Gutierrez 1150 (C.P.1613), Los Polvorines, Buenos Aires, Argentina and CONICET, Argentina\\
$^3$ Low Temperature Laboratory (OVLL), Aalto University, School of Science, P.O. Box 13500, 00076 Aalto, Finland \\
$^4$ Department of Applied Physics and COMP Center of Excellence, Aalto University School of Science, P.O. Box 11100,
00076 Aalto, Finland\\
$^5$ Institute for Complex Quantum Systems and Center for Integrated Quantum Science and Technology, Ulm University, Albert Einstein-Allee 11, 89069 Ulm, Germany}

\begin{abstract}
Work, moments of work and heat flux are studied for the generic case of a strongly driven two-level system immersed in a bosonic heat bath in domains of parameter space where perturbative treatments fail. This includes particularly the interplay between non-Markovian dynamics and moderate to strong external driving. Exact data are compared with predictions from weak coupling approaches. Further, the role of system-bath correlations in the initial thermal state and their impact on the heat flux are addressed. The relevance of these results for current experimental activities on solid state devices is discussed.
\end{abstract}
\pacs{03.65.Yz
05.70.Ln 
}

\maketitle
\section{Introduction}

The last years have seen a rapidly growing interest in thermodynamical properties of small systems, where fluctuations are essential and provide deeper insight in the changeover from microscopic to macroscopic behavior. Accordingly, concepts well-established in classical systems such as work and heat require a careful analysis for quantum mechanical aggregates \cite{Deffner2010,Hekking13,Solinas2013,Salm14}.
 The same is true for fluctuation relations such as the Jarzynski \cite{jarzynski:1997} or the Crooks \cite{crooks:1999} relation which on the macro level provide powerful tools to analyze situations far from equilibrium e.g.\ in biological and soft matter structures \cite{seifert2012}. Thus, on the micro level, a number of possible realizations including atomic systems \cite{rossnagel:2014,correra:2014} and mesoscopic solid state devices \cite{saira:2012} have been put forward to access signatures of quantum thermodynamics \cite{Engel2007, esposito:2009} and specific experiments are currently under preparation. Theory is now challenged to provide tools and methodologies to understand actual realizations.

 The problems encountered by theory are related to basically two issues, namely, the quantum measurement problem \cite{campisi:2011,Roncaglia2014} and the problem of describing dissipative quantum systems at very low temperature and in presence of also strong external time-dependent fields. This is a regime, where one expects, particularly for pulsed fields, a subtle interplay of non-Markovian dynamics and driving. With respect to the first topic, the two measurement protocol has been shown to provide at least formally a consistent basis for the detection of work and its moments \cite{campisi:2011}. Since work is not a decent quantum mechanical observable \cite{Talkner07}, it can only be defined "operationally" as the difference of eigen-energies before and after an external drive weighted by the thermal initial distribution and driving dependent transition probabilities. While this recipe can, at least in principle,  be implemented in an actual experiment for isolated systems, the situation for open (dissipative) systems is more intricate. Energy projective measurements of the full compound including the environmental degrees of freedom are not feasible, particularly when system degrees of freedom are strongly correlated or even entangled with those of thermal reservoirs. A prominent example that received much attention recently are reservoirs with sub-Ohmic mode distribution \cite{hur2007,kast:2013,Ank2014}.

\begin{figure}
  \centering
 \includegraphics[width=0.8\linewidth]{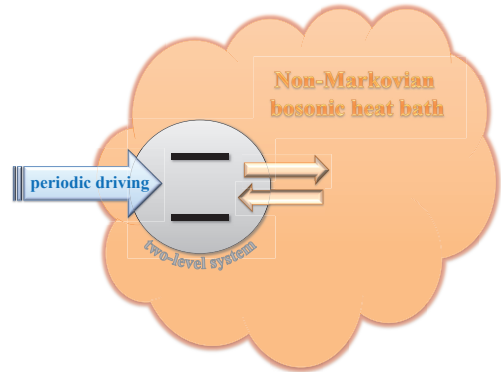}\\
  \caption{A two-level system is immersed in a bosonic heat bath and subject to an external time-dependent field over a finite period of time. The work exerted onto the compound partially leads to a change of internal energy and partially to heat flux into the bath. Intricate correlations between non-Markovian dynamics and stronger driving necessitate a non-perturbative treatment at very low temperatures.}\label{fig:theme}
\end{figure}

For the second topic, conventional approaches to capture the reduced dynamics of dissipative quantum systems comprise powerful methods such as master or Lindblad equations. However, these treatments are restricted to the domains of sufficiently weak system-bath interaction, sufficiently elevated temperatures (Markovian dynamics), and sufficiently weak driving (see e.g.\ \cite{breuer:2007}). Beyond, corresponding predictions become unreliable and non-perturbative formulations must be applied \cite{weiss2008}, for example, path integral Monte Carlo techniques, density matrix renormalization or stochastic Liouville-von Neumann equations. Interestingly, at least to our knowledge, detailed studies for work, its moments, and heat flux between a system of interest and thermal reservoirs in presence of moderate to strong driving (with respect to amplitude and driving frequency) and for low temperatures have not been performed yet. Here, we present first results to close this gap by providing benchmark data for the generic case of a dissipative two level system (cf. Fig.~\ref{fig:theme}).
Exact numerical simulations within the recently developed stochastic Liouville-von Neumann scheme (SLN) \cite{Stock2002,schmidt2011} are compared to perturbative ones obtained within a simple Lindblad type of master equation and a quantum jump treatment\cite{Hekking13}. Analytical calculations allow for a qualitative and in certain cases also quantitative understanding of the driven quantum dynamics. In addition, the SLN gives access to the impact of correlations between system and bath in thermal equilibrium. Namely, when one starts from an initially factorized state between system and reservoir which is the typical assumption in weak coupling treatments, a heat flux associated with these correlations is induced \cite{Ank2014}. We will further apply two different protocols to extract the work which are both based on the two measurement scheme but on different experimental realizations: One monitors the dynamics of a system observable (power operator), while the other monitors the energy transfer with the reservoir (photon emission/absorption processes).

The paper is organized as follows. After a discussion of the model, we give a brief account of the three methods to treat open system dynamics employed here (Sec.~\ref{sec2}). The first and the second moment of work are the subject of Sec.~\ref{sec3}, where we also present analytical findings. In Sec.~\ref{sec4} the heat flux is obtained within the exact SLN scheme including the role of initial correlations. The latter ones are further addressed in Sec.~\ref{sec5} before we conclude and give prospects for future developments in Sec.~\ref{sec6}.

\section{Quantum dynamics}\label{sec2}
We consider a two level system (TLS) subject to a time-dependent driving force, i.e.,
\begin{eqnarray}\label{hs}
H_{S}(t)&=&H_0+H_D(t)=- \frac{\hbar\Delta}{2}\sigma_{x}+\lambda(t)\,  \sigma_z\, ,
\end{eqnarray}
where
\begin{equation}\label{driving}
\lambda(t)=\lambda_0 \sin[\Omega (t-t_i)] \, \Theta_{if}(t)
\end{equation}
has a finite range with $\Theta_{if}(t)=1$ for $t\in [t_i, t_f]$ and zero elsewhere.

The TLS is immersed in a bosonic bath (spin-boson model \cite{weiss2008}) with Hamiltonian $H_R=\sum_k \hbar\omega_k b_k^\dagger b_k$, cf.~Fig.~\ref{fig:theme}, so that the Hamiltonian of the full compound has the standard form $H(t)=H_{S}(t)+H_{I}+H_{R}$ with $H_{I}=\sigma_z \mathcal{E}$, where $\mathcal{E}=\sum_k c_k (b_k^\dagger+b_k)$ is the bath force. In the continuum limit, the effective impact of the bath onto the system is then fully described by the spectral density of its modes $J(\omega)$ and its thermal energy $k_{\rm B}T\equiv 1/\beta$. In the remainder of this work, we focus on an Ohmic reservoir with large cut-off frequency $\omega_c$, i.e.,
 \begin{equation}\label{spectral}
 J(\omega)=\hbar \eta \omega\,  f_c(\omega/\omega_c)\, ,
 \end{equation}
 where $\eta$ is a dimensionless coupling constant and $f_c(x)$ a cut-off function with $f(x=0)=1, f_c(x\gg 1)\to 0$. Here, we choose $f_c(x)= 1/(1 + x^2)^2$ but the results shown below are not very sensitive to the specific form of the cut-off function due to $\omega_c\gg \Delta, \Omega$.

Now, given an initial density operator ${\cal W}(0)$ of the full compound, the reduced dynamics is determined by
\begin{equation}\label{reduced}
\rho(t)={\rm Tr}_R \{ U(t,0) {\cal W}(0) U(t,0)^\dagger\}\, ,
\end{equation}
with the time evolution operator $U(t,0)=\mathcal{T} \exp[-\frac{i}{\hbar} \int_0^t ds H(s)]$ and the trace performed over the environmental degrees of freedom only. At low temperatures, the dynamics of driven open quantum systems is a challenging task since bath induced memory effects (non-Markovian dynamics) are intermingled with driving induced transitions. Memory effects
appear on the time scale ${\rm max}\{\hbar\beta, 1/\omega_c\}$ which in the regime $\omega_c\hbar\beta\gg 1$ as considered here, grows with decreasing temperature. In case of periodic driving, the reduced system will approach a non-equilibrium steady state for longer times and displays transient behavior initially. In the context of work, one often  addresses only this latter time domain as external fields appear in form of pulses relatively short compared to time scales where the dynamics becomes stationary. Non-perturbative treatments are thus of paramount importance to arrive at quantitatively reliable predictions.
Here, we compare a numerically exact formulation, the Stochastic Liouville-Von Neumann equation (SLN), with two approximate approaches, namely, the Lindblad Master equation (LME) and the Quantum Jump method (QJ).

\subsection{Stochastic Liouville-von Neumann equation}
The SLN can be directly derived from the exact Feynman-Vernon path integral formulation \cite{weiss2008} for the reduced density operator (\ref{reduced}). An unraveling procedure then leads to the SLN \cite{Stock2002,Stock2004} which for the driven spin-boson model acquires the form
\begin{equation}\label{sln}
\nrl{\dot{\rho}}(t)= -\ihq [H_{S}(t),\nrl{\rho}]+\ihq \xi(t) [\sigma_{z},\nrl{\rho}]+\frac{i}{2}\nu(t)\{\sigma_{z},\nrl{\rho}\}\, .
\end{equation}
This equation holds for a single noise realisation $\mathcal{Z}\equiv\{\xi,\nu\}$, whereas the physical reduced density $\rho(t)$ is gained by averaging over a sufficiently large number of noise realisations, i.e.,  $\rho(t) = \sEW{\nrl{\rho}(t)}$. While (\ref{sln}) is local in time, yet, the full non-Markovian dynamics is captured in $\rho(t)$. The correlation functions of the two complex-valued noise forces $\xi(t)$ and $\nu(t)$ reproduce the complex-valued and non-local in time force autocorrelation function of the bath.
Since effectively the noise forces appear as driving forces, an additional external driving is easily taken into account in (\ref{sln}) for arbitrary driving strengths and driving frequencies.

 The initial state on which this SLN is based, is a factorizing state $\mathcal{W}(0)=\rho(0) \otimes\exp(-\beta H_R)/Z_R$ with the initial density $\rho(0)$ of the TLS and the partition function $Z_R$ of the reservoir. This allows for a direct comparison with the approximate formulations for which this initial state is always taken for granted. However, the impact of correlated initial states can be explored as well within the SLN as will be discussed below.

\subsection{Lindblad master equation}
A very powerful instrument to simulate the dynamics of open quantum systems are LME.  While originally formulated within the mathematical theory of semi-groups, they can be derived from  system+reservoir models by employing Born-Markov perturbation theory together with a coarse graining procedure in time, see e.g.\ \cite{Carmichael99,breuer:2007}. For driven systems, one further has to impose weak driving (small amplitude and/or slow driving).

For the system under consideration  (\ref{hs}), one has
\begin{eqnarray}
\dot{\rho} &=& -\frac{i}{\hbar} \left[H_S(t),  {\rho} \right] \nonumber\\
 &&+  \sum_{k=0}^1 \left( {L}_{k} {\rho}  {L}_{k}^\dagger - \frac{1}{2}  \left\lbrace {L}_{k}^\dagger {L}_{k}   ,{\rho} \right\rbrace \right),
\label{eq:LME}
\end{eqnarray}
with Lindblad operators $L_{0}=\sqrt{\gamma_{0,1}} |0\rangle \langle 1|$ and $L_{1}=\sqrt{\gamma_{1,0}} |1\rangle \langle 0|$, where $|0\rangle$, $|1\rangle$ being the eigenstates of $H_0$ with values $\mp \hbar \Delta /2$, respectively.

 The transition rates $\gamma_{nk}$ take the usual form
 \begin{eqnarray}\label{born-rates}
 \gamma_{0, 1} &= &\frac{\eta}{2}\Delta \left[1+{\rm coth}(\Delta\hbar\beta/2)\right],\ \gamma_{1, 0}=\gamma_{0, 1} {\rm e}^{-\Delta\hbar\beta}
 \end{eqnarray}
 with coupling constant $\eta$ as in (\ref{spectral}). This and extended schemes working e.g.\ in a Floquet representation have been recently applied in the context of work and its distribution for open quantum systems \cite{silaev2014,gasparinetti:2014,Lang2014,Cuetara2014}.

\subsection{Quantum jump method}

The QJ has been pioneered in quantum optics to describe emission and absorption processes of single photons by few level systems (atoms) \cite{dalibard1992}. In more general terms, the method exploits the probabilistic nature of the quantum mechanical time evolution by constructing the dynamics $|\psi(t)\rangle \to |\psi(t+\Delta t)\rangle$ over a time interval $\Delta t$ according to sequences of jumps between energy levels with transition probabilities determined by the corresponding Hamiltonian \cite{plenio1998,Carmichael99}. Practically, one uses a Monte Carlo procedure to sample individual jump trajectories and expectation values are obtained by averaging over a sufficiently large number of realizations.

This method has recently been formulated to obtain the work of driven TLS interacting with bosonic baths \cite{Hekking13}. In this context, one works with a system-bath coupling in rotating wave approximation, i.e.,
\begin{equation}\label{hirwa}
H_I^{\rm RWA}= \sum_k \left( c_k\,  b_k \sigma_+ + c_k^*\, b^\dagger_k \sigma_- \right)
\end{equation}
with spin raising/lowering operators $\sigma_\pm=\sigma_x\pm i\sigma_y$. Accordingly, the system-bath interaction captures the exchange of on-shell photons which can be easily recorded numerically to obtain the heat transfer from/into the bath during the time evolution.
The corresponding absorption/emission rates of the TLS are obtained from a Born-Markov treatment, cf.~(\ref{born-rates}). The change in system energy is monitored by recording the last photon exchange before and the first after the drive which implies two times projective measurements as shown in \cite{Hekking13}. This then allows very effectively to extract the work and its distribution for an open few level quantum system \cite{Hekking13}. As an approximate method, this QJ applies for weak system-bath couplings and weak driving similar to the LME. In fact, one can prove that formally the QJ approach leads to the LME.

Below we will use the LME and the QJ  to obtain the work according to two different schemes which correspond to two different experimental situations: The LME describes the dynamics of a system observable, the power operator, while the QJ monitors the energy exchange with the reservoir, i.e. photon emission/absorption processes. Thus, both methods provide identical results in the regime, where these schemes are expected to be reliable, but may differ beyond.

\section{Moments of work}\label{sec3}
Since work itself is not a proper quantum observable, the calculation of its moments must be performed with care \cite{campisi:2011}.
A consistent formulation has been provided by the two measurement protocol (TMP) \cite{Talkner07} which even allows to retrieve the full distribution of work \cite{esposito:2009,campisi:2011}. According to this scheme, the probability to measure energy $E_i$ at time $t=t_i$ and $E_f$ at time $t_f$  is given by
\begin{equation}
P[E_f, E_i] ={\rm Tr}\{ \Pi_{E_f}\, U(t,0)\, \Pi_{E_i} \mathcal{W}(0)\, \Pi_{E_i} \, U^\dagger(t,0)\, \Pi_{E_f}\}\, ,
\end{equation}
where $\Pi_{E_{i/f}}=|E_{i/f}\rangle \langle E_{i/f}|$ are projection operators on energy eigenstates at $t=t_i$ and $t=t_f$, respectively. The work distribution then follows from $p(W)=\sum_{E_i, E_f} \delta[W-(E_f-E_i)] P[E_f, E_i]$. One can easily show \cite{Solinas2013,suomela:2014} that the first two moments of work derived from this distribution can be expressed in terms of the power operator \cite{Solinas2013}
\begin{equation}
  {P}_W=\frac{\partial H_S}{\partial t}\equiv\frac{\partial H_S}{\partial \lambda}\, \dot{\lambda}(t)
\end{equation}
as
\begin{eqnarray}\label{moments}
\langle W\rangle_t &=&\int_0^t ds\ \langle P_W^{H}(s)\rangle\nonumber\\
\langle W^2\rangle_t &=& \int_0^t ds \int_0^t du\ \langle P_W^{H}(s)P_W^{H}(u)\rangle\nonumber\\
&=& 2 \int_0^t ds \int_0^s du\ {\rm Re}\{\langle P_W^H(s)P_W^H(u)\rangle\}\,
\end{eqnarray}
 if expectation values are taken with respect to $\sum_{E_i} \Pi_{E_i} \mathcal{W}(0)\, \Pi_{E_i}$. Here, $P_W^H(t)$ denotes the Heisenberg operator to $P_W$. These expressions are particularly convenient for quantum open systems for which a diagonalization of the full Hamiltonian is out of reach. Instead, one has to calculate time-dependent moments of system observables which can be achieved based on the methods described above for the reduced density operator with properly chosen initial states. In principle, higher moments can be calculated as well, however, corresponding results are not consistent with those obtained within the two measurement protocol, see e.g.\ \cite{suomela:2014}.

In the context of work, the initial state is typically a thermal state. According to weak coupling approaches such as LME and QJ, one writes $\mathcal{W}(0)= \rho(0)\otimes {\rm e}^{-\beta H_R}/Z_R$ with $\rho(0)={\rm e}^{-\beta H_S(0)}/Z_0$ and where $Z_{R/0}$ are the partition functions of the bare system and bath, respectively. This initial density is diagonal in the basis of factorized energy eigenstates of system and bath. In case of a TLS one has
\begin{equation}\label{simpleequi}
 \rho(0)=\frac{1}{Z_0} \left(e^{\frac{\hbar\Delta\beta}{2}}|0\rangle \langle 0|+e^{-\frac{\hbar\Delta\beta}{2}}|1\rangle \langle 1|\right)\, \label{eq:ini}
 \end{equation}
 with $Z_0=2\cosh(\frac{\hbar\Delta\beta}{2})$ and $|0\rangle$, $|1\rangle$ being eigenstates of $H_0$ with eigenvalues $\mp \hbar\Delta/2$, respectively.

 However, for any finite coupling the true thermal state is a correlated state of TLS and bath, i.e.\ ${\cal W}_\beta={\rm e}^{-\beta H(0)}/Z$, and the corresponding reduced distribution $\rho_\beta={\rm Tr}_R\{{\cal W}_\beta\}$ is not of Gibbs form \cite{lutz2009,Ank2014}. Accordingly, in actual experiments the true initial state may be only of the form (\ref{eq:ini}) for extremely weak coupling, an issue, that will be addressed in more detail below.

For the simulations performed in the sequel, we use natural units, i.e. $\Delta=1$, $\hbar =1$ and $m=1$, restrict ourselves to the resonant situation $\Omega=\Delta$, and consider $\lambda_0\geq 0$. The bath cut-off is taken as $\omega_c=10$ for the SLN simulations. Further, for the drive we set $t_i=0$ and $t_f=3 \pi$ as not indicated otherwise.

\subsection{First moment}
According to (\ref{hs}) and (\ref{moments}) we start with
\begin{equation}\label{firstmoment}
\EW{W(t)}=\int\limits_{0}^{t} ds\ \dot{\lambda}(s)\EW{\sigma_{z}(s)}\, \label{eq:work}
\end{equation}
and analyze its dependence on driving strength and temperature. Apart from numerical data, transparent analytical expression are available for negligible system-bath interaction.

\subsubsection{Numerical results}
\begin{figure}
\includegraphics[width=0.6\linewidth]{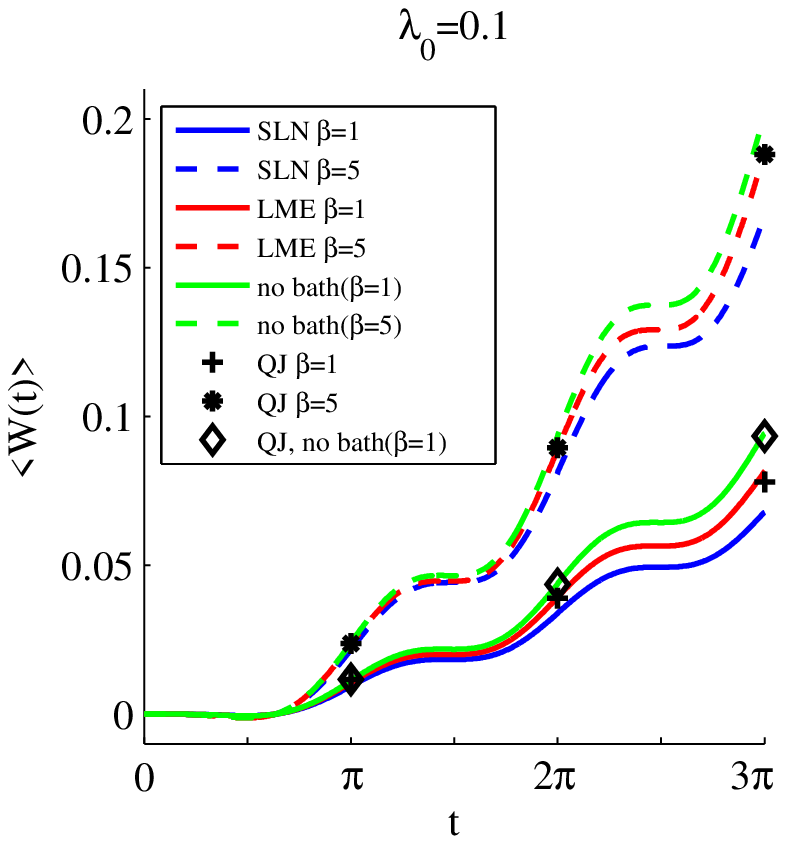}
\includegraphics[width=0.6\linewidth]{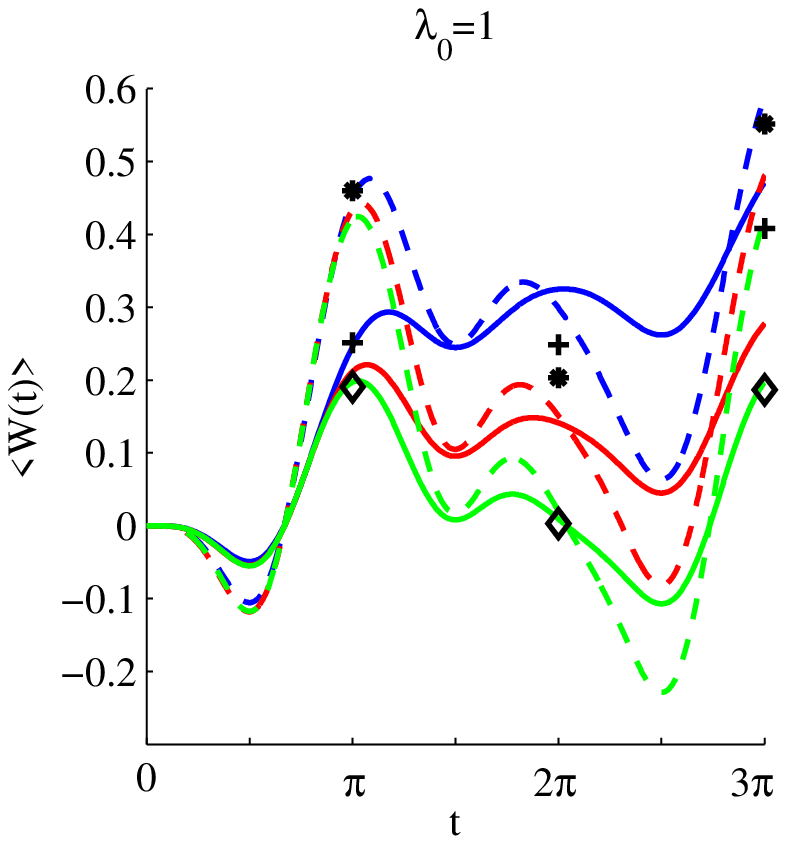}
\includegraphics[width=0.6\linewidth]{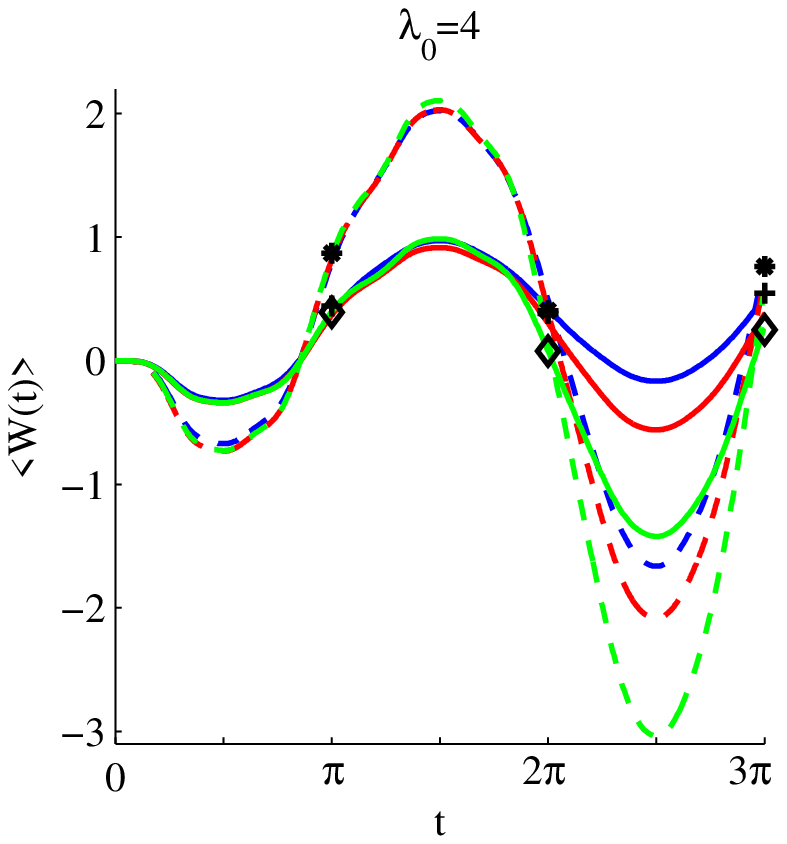}
\caption{\label{fig:work} Work according to the SLN (blue), the QJ (markers), and the LME (red) approach
for various driving amplitudes $\lambda_0$ and inverse thermal energies $\beta$ in natural units. All data are for the same coupling strength $\eta = 0.05$ and the driving force acts in the time interval $[0, 3\pi]$. Shown are also data in absence of a dissipative bath (green) with an initial thermal state at $\beta=1$ (solid) as well as $\beta=5$ (dashed).}
\end{figure}
Figure \ref{fig:work} displays data for all three approaches with data points of the QJ only included for multiples of $\pi$ for clarity. For comparison, data of the bare TLS (i.e. without bath) with a thermal initial state (\ref{simpleequi}) are shown as well. In this latter case, all approaches, SLN, LME, and QJ, provide identical results, of course.

For finite but weak system-bath coupling and in the regime of weak driving ($\lambda_0=0.1$, upper panel), the work is an increasing function of time and the approximate methods LME and QJ reproduce the exact SLN data quite accurately even at low temperatures $\beta=5$. As expected, LME and QJ produce identical data within statistical errors.
  Apparently, $\langle W(t)\rangle$ is smaller for higher temperatures since then initially the population difference between the two eigenstates states is smaller compared to the low temperature situation. Further, a finite system-bath coupling reduces the work compared to the bare dynamics since heat is transferred to the reservoir as well.

 The situation is substantially different for stronger driving. For $\lambda_0=1$, the work for  finite system-bath coupling is, after an initial transient, always positive and always exceeds that of the bare system. The approximate approaches provide these features qualitatively, however, quantitatively they differ quite substantially from the exact results.  We note that extended scheme for time-dependent driving with the LME \cite{silaev2014,gasparinetti:2014,Lang2014,Cuetara2014} and slow driving with the QJ \cite{Suomela2014b} have been developed recently, which account for the influence of the driving onto the dissipator in a more elaborate way, e.g.\ using the Floquet formalism. Instead the SLN applies to arbitrary pulse forms and driving strengths, particularly to those obtained from optimal control schemes \cite{schmidt2011}.
 For even stronger driving $\lambda_0=4$, the dynamics tends to be dominated by the system dynamics with the impact of the bath playing 
only a minor role. Accordingly, the agreement between the three numerical methods improves again.
We note that due to the different quantitites on which their work calculation is based, LME and QJ predictions differ mostly in the 
regime, where driving amplitude, driving frequency, and system energies are of the same order of magnitude. This domain is beyond their range of validity and they thus produce uncontrolled errors.

\subsubsection{Analytic results}\label{analytical:first}
For weak driving and at or close to resonance $\Delta=\Omega$, one more conveniently starts from the rotated TLS in
a rotating wave approximation so that
\begin{equation}
H_S'(t)\approx H'_{RWA}(t)=\frac{\hbar\Delta}{2} \sigma_z +i \frac{\lambda_0}{2}\left( {\rm e}^{-i \Omega t}\sigma_++{\rm e}^{i \Omega t}\sigma_-\right)\, .
\end{equation}
Then, assuming negligible system-bath interaction, a simple calculation provides an explicit expression for $\langle \sigma_z(t)\rangle$. The result for the work according to (\ref{firstmoment}) becomes particularly transparent at times $t=N\pi, N=1, 2, 3, \ldots$
\begin{equation}
\frac{\langle W\rangle_N}{\hbar\Omega}\approx (2 P_g-1) \sin^2\left(\frac{N \pi \bar{\lambda}_0}{2}\right)  \left(1-\frac{\bar{\lambda}_0^2}{4-\bar{\lambda}_0^2}\right)\label{eq:waw}
\end{equation}
with $\bar{\lambda}_0=\lambda_0/\hbar\Omega$ and $P_g$ the initial population of the ground state $|0\rangle$ according to (\ref{eq:ini}).
This result describes the numerical data for the bare dynamics quite accurately. One observes that the work exerted onto the  TLS is for weak driving only limited by the initial ground state population and depends sinusodially on driving period and strength.

In the opposite regime of very strong driving, a perturbative treatment starts from (\ref{hs}) with $t_i=0$ for convenience, i.e.\
$H_S(t)= -(\hbar\Delta/2)\sigma_x + \lambda_0 \sin(\Omega t) \sigma_z$.
For $\lambda_0\gg \hbar\Delta$ transitions between diabatic states (eigenstates of $\sigma_z$) only occur close to $\Omega t=k \pi, k=1,2,3,\ldots$, and the drive sweeps very fast (with velocity $\lambda_0\Omega$) through the Landau-Zener region. Hence, a dressed tunneling picture applies with a polaron-transformed Hamiltonian
\begin{eqnarray}
\tilde{H}_S(t) &=& \frac{\hbar\Delta}{2}\left[ {\rm e}^{i\phi(t)}\sigma_++{\rm e}^{-i\phi(t)}\sigma_-\right]\, ,
\end{eqnarray}
where $\phi(t)=-\bar{\lambda}_0 \cos(\Omega t)$. The exponential ${\rm e}^{-i\phi(t)}=\sum_n (-i)^n J_n(\bar{\lambda}_0) {\rm e}^{-i n\Omega t}$ is dominated by the time independent part for $n=0$ so that we arrive at $\tilde{H}_{RWA}=\frac{\hbar\Delta_0}{2} \sigma_x$ with a dressed tunneling amplitude $\Delta_0=\Delta J_0(\bar{\lambda})$. This then implies for $t=N\pi, N=1, 2, 3, \ldots$
\begin{equation}
\frac{\langle W\rangle_N}{\hbar\Omega}= (2 P_g-1) \frac{\bar{\lambda}_0 J_0(\bar{\lambda}_0)}{1-J_0(\bar{\lambda}_0)^2} \cos(2\pi N) \sin[2\pi N J_0(\bar{\lambda}_0)]\, .\label{eq:was}
\end{equation}
While due to missing higher harmonics, this prediction for the work cannot be used for a quantitative comparison with the numerical data, it provides an at least qualitatively correct description with the correct order of magnitude of the oscillatory features. In the limit of very strong driving, the maximal work becomes for fixed $N\ll \sqrt{\bar{\lambda}_0}$,  and with $J_0(|x|)\approx \sqrt{2/\pi|x|}\cos(|x|-\frac{\pi}{4})$ for $|x|\gg 1$  independent of the driving amplitude such that
\begin{equation}
\left.\frac{\left|\langle W\rangle_N\right|}{\hbar\Omega}\right|_{\bar{\lambda}_0\gg 1}\leq (2 P_g-1) 4 N \, .\label{eq:was}
\end{equation}
This feature is directly reflects the energy saturation in a TLS. For long driving times $N\gtrsim \sqrt{\bar{\lambda}_0}\gg 1$, the work oscillates with an amplitude of order $(2 P_g-1) \sqrt{\bar{\lambda}_0}$ as also seen in the lower panel of Fig.~\ref{fig:work}.

\subsection{Second moment}
\begin{figure}
\includegraphics[width=0.45\linewidth]{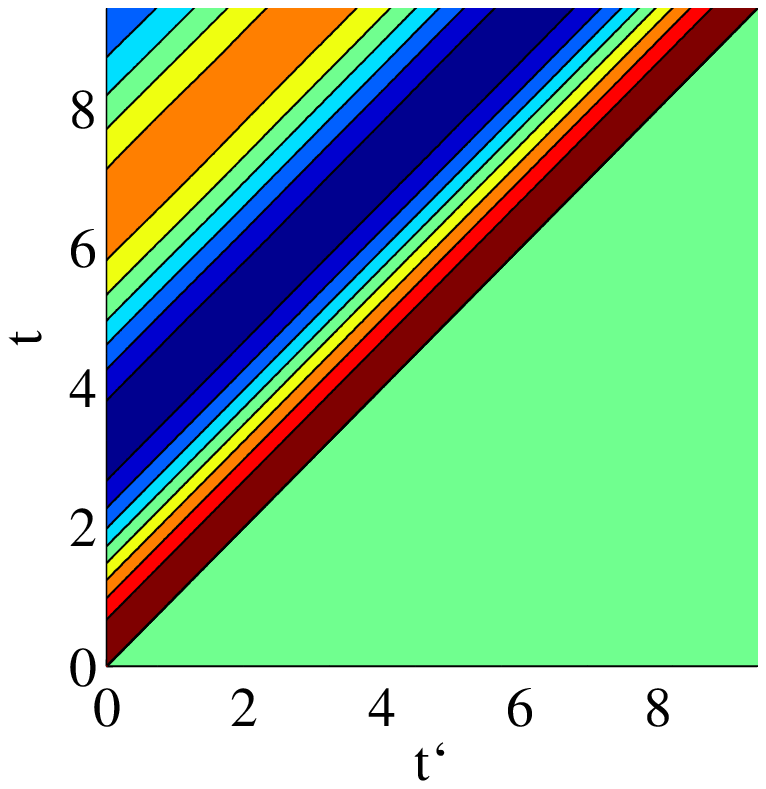}
\includegraphics[width=0.45\linewidth]{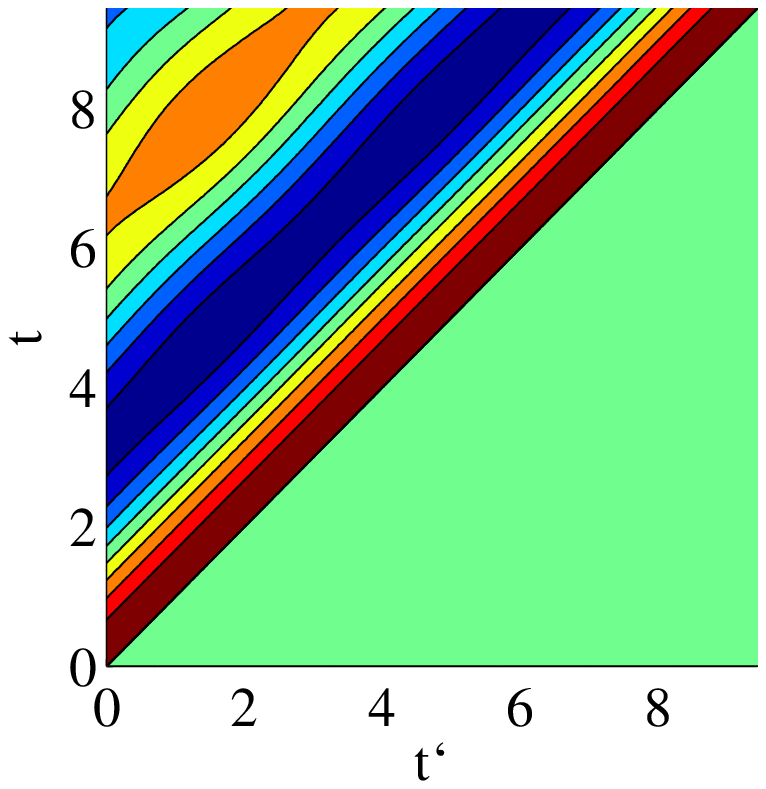}\\
\includegraphics[width=0.45\linewidth]{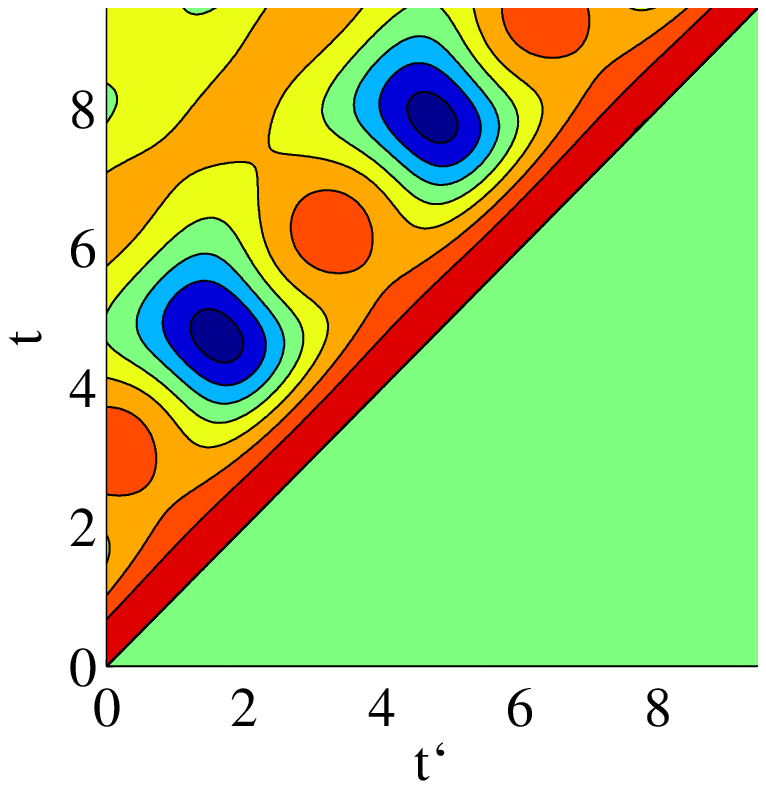}
\includegraphics[width=0.45\linewidth]{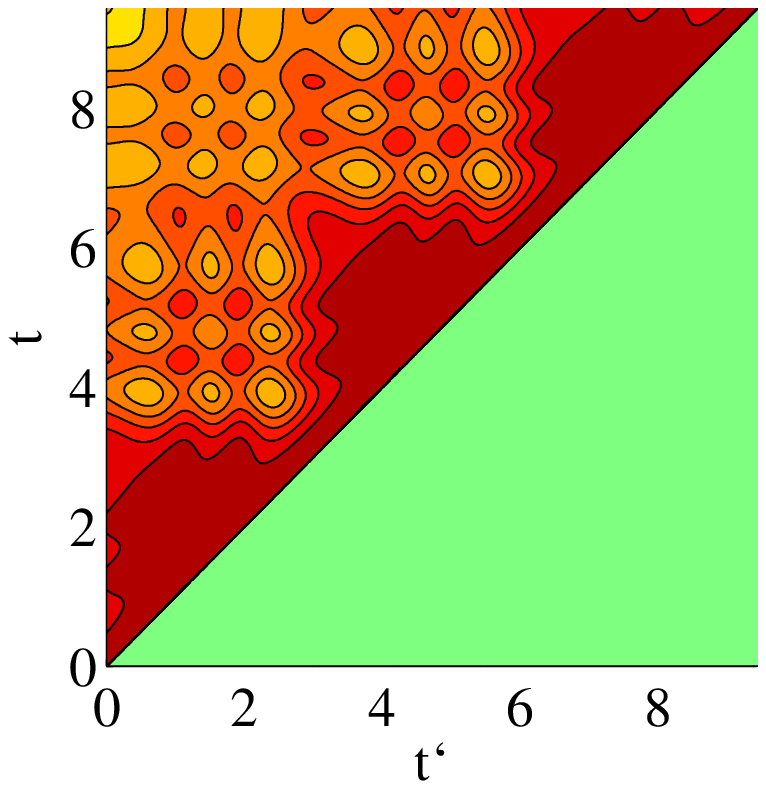} \caption{Contour plots of the real part of the two-time correlator ${\EW{\sigma_{z}(t)\sigma_{z}(t')}}$ for different driving amplitudes. Due to symmetry only the domain $t\geq t'$ is shown.
Upper left $\lambda_{0} = 0$, upper right $\lambda_{0} = 0.1$, lower left $\lambda_{0} = 1$ and lower right $\lambda_{0} = 4$. Other parameters are as in Fig.~\ref{fig:work}. Coulour scale goes from red (+1) to blue (-1), with green indicating zero.}  \label{fig:ttc}
\end{figure}
Along the lines of (\ref{moments}) we have for the second moment
\begin{equation}
\EW{W^2(t)}=2\int\limits_{0}^{t} ds \ \int\limits_{0}^{s} du\  \dot{\lambda}(s)\dot{\lambda}(u){\rm Re}\left\{\EW{\sigma_{z}(s)\sigma_{z}(u)}\right\}\label{eq:wsq}\,
\end{equation}
which includes the time-ordered two-time correlator
\begin{equation}
{\EW{\sigma_{z}(s)\sigma_{z}(u)}}={\rm Tr}\left\{ \sigma_{z}U(s,u) \sigma_{z} {\cal W}(u)U^{\dagger}(s,u)\right\}\, .
\end{equation}

\subsubsection{Numerical results}
The $\sigma_z$ correlator is an interesting dynamical quantity in itself as illustrated in Fig.~\ref{fig:ttc}. For increasing driving strength,  pronounced patterns with increasing fine structure reveal the underlying Floquet-modes of the dynamics also at finite dissipation. For the strongest driving, $\lambda_{0} = 4$, the real part of the two-time correlator ${\rm Re}\{\EW{\sigma_{z}(s)\sigma_{z}(u)}\}$ which enters the expression (\ref{eq:wsq}) is always positive.

\begin{figure}
\includegraphics[width=0.6\linewidth]{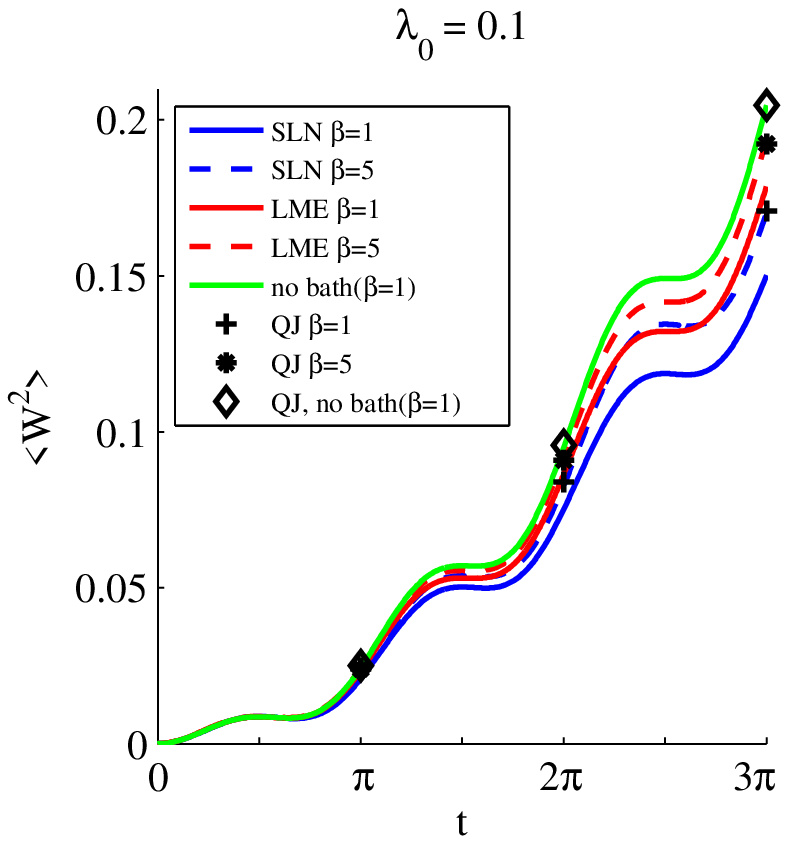}
\includegraphics[width=0.6\linewidth]{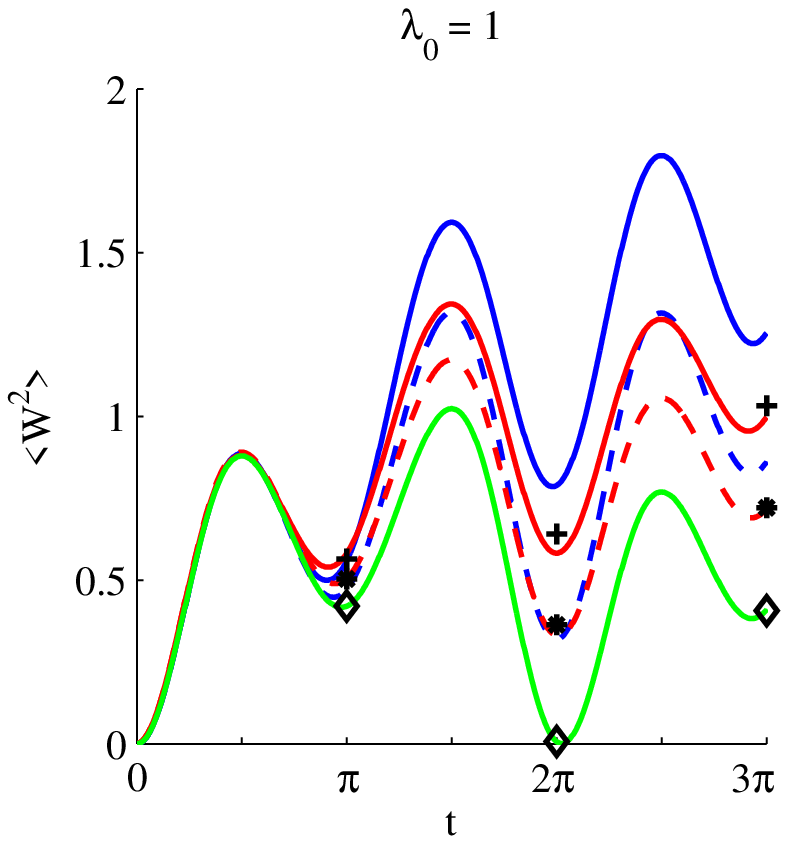}
\includegraphics[width=0.6\linewidth]{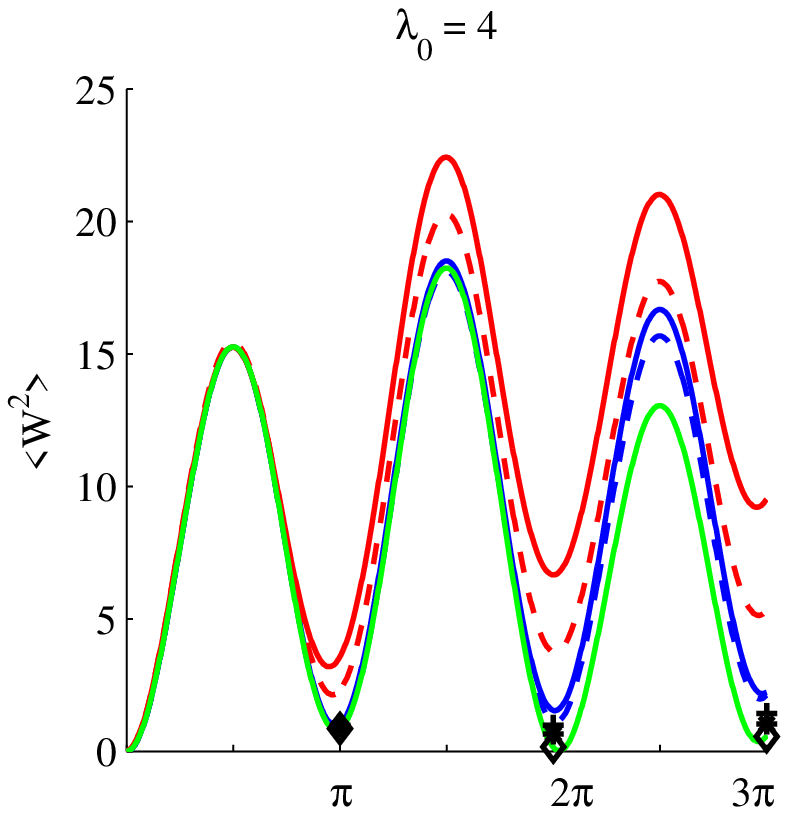}
\caption{$\EW{W^2 (t)}$ according to the SLN (blue), the QJ (markers), and the LME (red) approach for various driving amplitudes $\lambda_0$ and inverse thermal energies $\beta$ in natural units. Data in absence of a dissipative medium with an initial thermal state at $\beta=1$ are also shown. Other parameter are as in Fig.~\ref{fig:work}.}  \label{fig:wsq}
\end{figure}

   Data for the corresponding $\EW{W^2(t)}$ are depicted in Fig.~\ref{fig:wsq}. For weak driving, $\lambda_{0} = 0.1$, one observes a nearly monotonic behavior with superposed weak oscillations.  Exact results are accurately reproduced by both LME and QJ. For stronger driving deviations from the SLN results become more prominent and also deviations between LME and QJ data.  The tendency that dissipation reduces work fluctuations changes for stronger driving ($\lambda_0=1$), where 
fluctuations for finite dissipation are bounded from below from those in absence of a heat bath. As the driving becomes again the dominant feature, work fluctuations display oscillatory behavior. 
  Further, at least for weak to moderate driving one has $\EW{W^2 (t)} \sim \lambda_0^2$, a dependence verified by analytical results presented in the next section.
%
%
\subsubsection{Analytic results}
In the weak driving case, one proceeds as described in Sec.~\ref{analytical:first} and obtains
\begin{equation}
\frac{\langle W^2\rangle_N}{(\hbar\Omega)^2}= \sin^2(N\pi \bar{\lambda}_0/2)\, .
\end{equation}
The fluctuations around the mean value are thus independent of the initial population and limited by the level splitting of the TLS. This result is in accurate agreement with the numerical data with a quadratic dependence on the driving amplitude as long as $N\bar{\lambda}_0\ll 1$.

Likewise, for strong driving, we arrive with $\langle \sigma_z(s)\sigma_z(u)\rangle\approx \cos[\Delta_0(s-u)]$ at
 \begin{equation}
\frac{\langle W^2\rangle_N}{(\hbar\Omega)^2}= 2\bar{\lambda}_0^2\,  \frac{  J_0(\bar{\lambda}_0)^2}{[1- J_0(\bar{\lambda}_0)^2]^2}\ \left\{1-(-1)^N \cos[ N\pi J_0(\bar{\lambda}_0)]\right\}\,
\end{equation}
which follows a quadratic dependence on the driving amplitude as long as it is not extremely large. In this latter case, for fixed $N\ll \sqrt{\bar{\lambda}_0}$ and in leading order we have
 \begin{equation}\label{w2strong}
\left.\frac{\langle W^2\rangle_N}{(\hbar\Omega)^2}\right|_{\bar{\lambda}_0\gg 1}\approx
\left\{
\begin{array}{ll}
N^2 \cos^4(\bar{\lambda}_0-\frac{\pi}{4}) &\ , \ \ N\, {\rm even}\\
\frac{8 \bar{\lambda}_0}{\pi}\cos^2(\bar{\lambda}_0-\frac{\pi}{4})&\ , \ \ N\, {\rm odd}
\end{array}\right.\, .
\end{equation}
  Hence, at odd multiples of half a Rabi cycle fluctuations grow with the driving amplitude, while at multiples of full cycles they grow with the total driving time. Numerical data at a moderate driving amplitude $\bar{\lambda}_0=4$, depicted in Fig.~\ref{fig:wsq}, are well described by the full time dependent $\langle W^2\rangle_t$  for $t\lesssim 3 \pi/2$ and are thus in agreement with (\ref{w2strong}) for $N<2$. For larger times, exact fluctuations further increase with  superposed oscillations of order $\bar{\lambda}_0$.

\section{Heat flux}\label{sec4}

The heat flux between system and reservoir is an important measure which together with the change in system energy allows to determine the work. In fact, current experimental activities in solid state devices exploiting fast thermometry aim at exactly this \cite{cryo2,cryo3}. Within the system+reservoir model, one derives an explicit expression based on
$\frac{d}{dt}\langle H_S^H(t)+H_I^H(t)+H_R^H(t)\rangle = \partial \langle H_D^H(t)\rangle / \partial t$ with the superscript $H$ denoting the corresponding Heisenberg operators. Treating then the terms on the LHS separately, one obtains the first law of thermodynamics, i.e.,
\begin{eqnarray}
  \EW{W}_t&=&\int_0^{t} du \EW{\frac{\partial H_{D}^H(u)}{\partial u}} \nonumber\\
  &=& \Delta E(t) +\ihq \int_0^{t} du \EW{\left[H_0^H(u),H_I^H(u)\right]}\,  \label{firstlaw}
\end{eqnarray}
with $\Delta E(t)=\EW{H_S^H(t)}-\EW{H_0(0)}$. Here,  we took into account that $H_D(0)=0$ and $[H_D, H_I]=0$ [cf.~(\ref{hs})]. The integrand in the last part is the heat flux $j_Q(t)$ and its time integral is the total heat $Q(t)$ exchanged during the time interval $t$.

Now, the SLN dynamics respect the first law as well, of course. Accordingly, one easily derives  from (\ref{sln})  an equation for $d\langle H_S^H(t)\rangle/dt$ which, when integrated over time and compared with (\ref{firstlaw}), provides an explicit expression for the heat flux in terms of the SLN formulation, namely,
\begin{equation}\label{heatflux}
j_{Q}(t)=-\Delta\ \sEW{\xi(t)\EW{\sigma_y (t)}}\, .
\end{equation}
Note that this line of reasoning only applies on the level of expectation values averaged over noise realizations, not on the operator level. In (\ref{firstlaw}) the operator of the heat flux acts in full Hilbert space, $(i/\hbar)[H_0^H(u),H_I^H(u)]=-\Delta \sigma_y \mathcal{E}$ with the bath force operator $\mathcal{E}$ as introduced below (\ref{driving}). In contrast, in  (\ref{heatflux}) the noise force $\xi(t)$ is a complex-valued number which for a single realization has no physical meaning. Anyway, based on (\ref{heatflux}) it is now straightforward to gain within the SLN approach numerically exact data for $j_Q$ and $Q$.\footnote{In the weak coupling regime analytical expressions for the heat exchange have recently been derived in\cite{Carrega2014}.}

%
\begin{figure}[h!]
\includegraphics[width=0.8\linewidth]{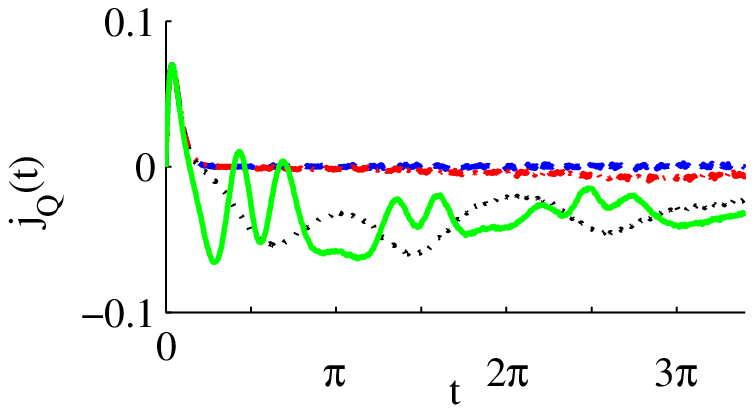}
\includegraphics[width=0.8\linewidth]{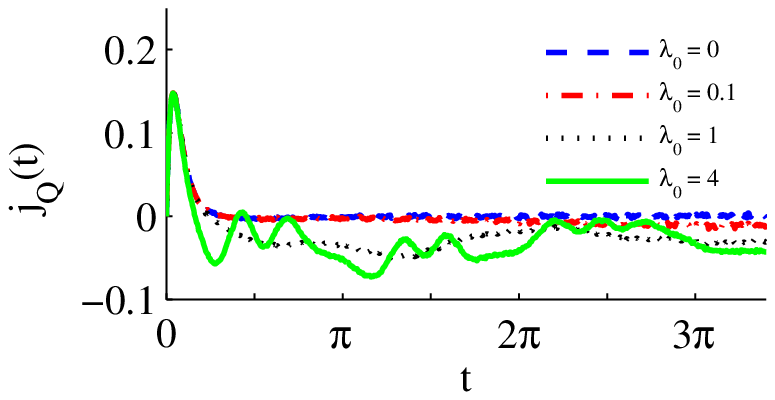}
\caption{Heat flux $j_{Q}$ for different driving amplitudes and for $\beta = 1$ (top) and $\beta = 5$ (bottom) for an initially factorizing state. Other parameters are as in Fig.~\ref{fig:work}.}\label{fig:hf1}
\end{figure}
%
Figure \ref{fig:hf1} shows the heat flux $j_{Q}$ for different driving amplitudes and two different bath temperatures. A striking feature is the peak in the heat flux at early times which is even present if there is no driving at all (cf.~\cite{Ank2014}). We attribute it to the factorizing initial condition (\ref{eq:ini}): The dynamics according to the full Hamiltonian immediately tends to correlate bath and system which in turn is related to heat exchange. This initial heat transfer grows at lower temperatures and then becomes even larger in amplitude as the heat flux due to an external drive. More details will be addressed in the next section.
Apart from that, for stronger driving, the heat flow starts to oscillate and may even revert its direction. Note that in the time interval displayed in Fig.~\ref{fig:hf1}, a stationary state with a strict periodic time dependence has not approached yet.

\begin{figure}[h!]
\includegraphics[width=0.8\linewidth]{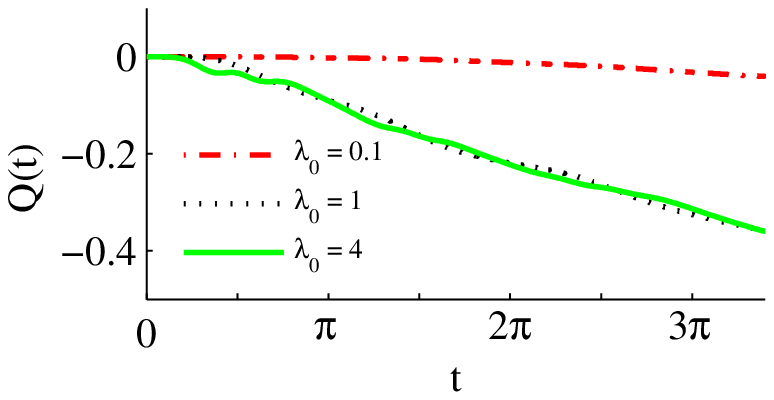}
\includegraphics[width=0.8\linewidth]{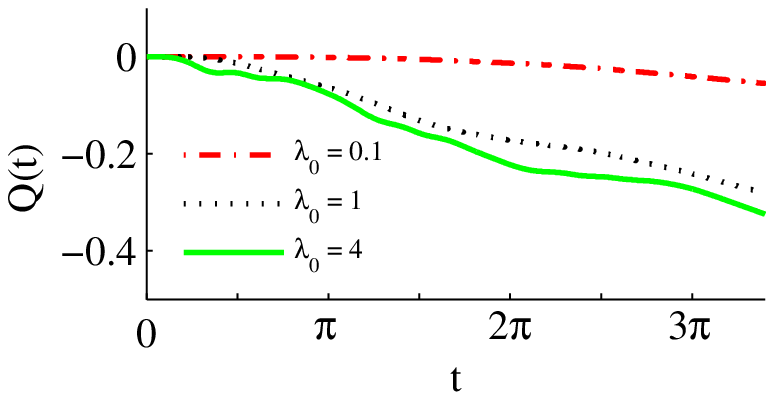}
\caption{As in Fig.~\ref{fig:hf1} but for the heat $Q$. The heat flux arising from the factorizing initial conditions (initial peak in Fig.~\ref{fig:hf1}) has been subtracted, see text for details.}\label{fig:q1}
\end{figure}

Based on the heat flow, one can also calculate the heat as depicted in Fig.~\ref{fig:q1}. There, the heat related to the initial peak in the heat flux is subtracted for better comparison. For a given set of parameters, the net heat transferred to the bath ($Q<0$) seems to saturate with further increasing driving amplitude. Namely, according to $Q=\langle W\rangle-\Delta E$ the exchanged heat is limited by the maximal change in internal energy $|\Delta E|\leq \Delta_0 \hbar$ and the maximal work specified for strong driving in (\ref{eq:was}). For  $\bar{\lambda}_0\gg 1$ one thus has $|Q|/\hbar\Omega \leq (2 P_g-1) 4 N$ independent of the driving amplitude.

%
%

\section{Factorizing initial conditions}\label{sec5}
\begin{figure}[h!]
\includegraphics[width=0.48\linewidth]{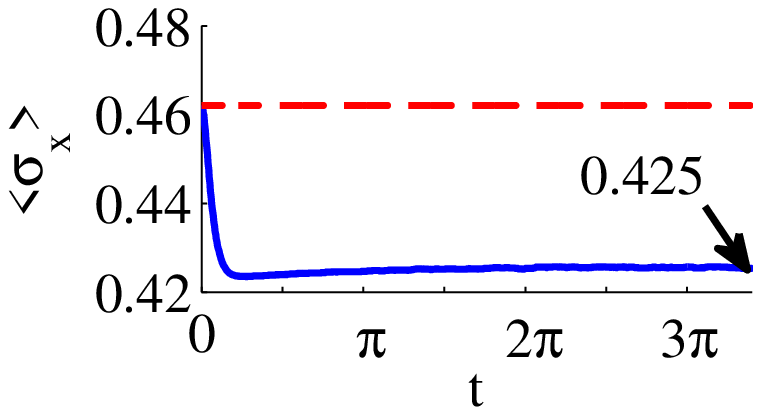}
\includegraphics[width=0.48\linewidth]{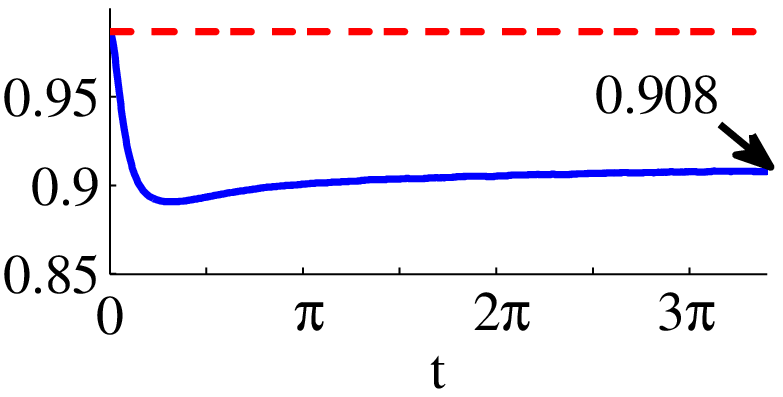}
\caption{Thermalisation of $\EW{\sigma_{x}}$ according to the SLN starting from an initially factorizing state (\ref{factorize}) with $\rho(0)$ as in (\ref{eq:ini}) in absence of external driving: $\beta = 1$ (left) and $\beta = 5$ (right). For comparison the bare thermal expectation values are depicted as well (dashed, red). Other parameters are as in Fig.~\ref{fig:work}.}\label{fig:therm}
\end{figure}

\begin{figure}[t]
\includegraphics[width=0.7\linewidth]{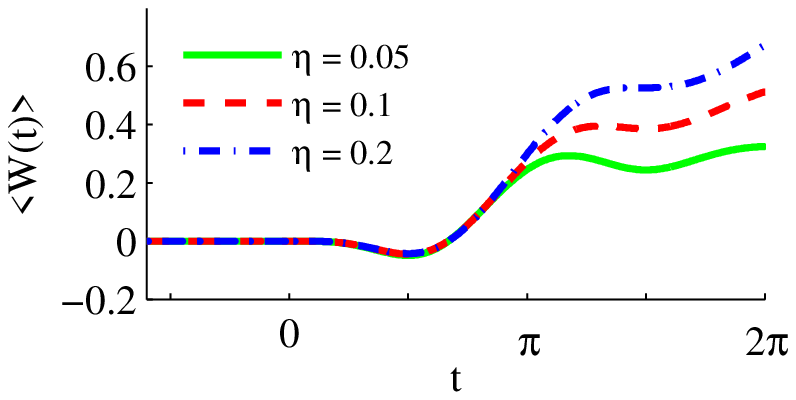}
\includegraphics[width=0.7\linewidth]{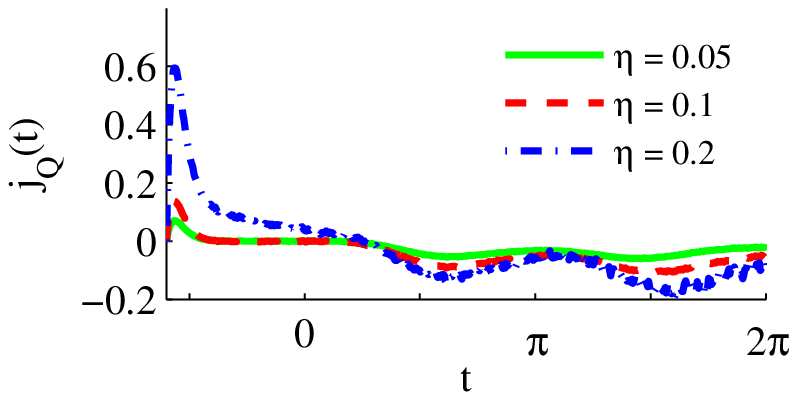}
\caption{\label{fig:delayed} Work and heat flux with delayed driving for $\beta=1, \lambda_0=1$ and various coupling parameters.}
\end{figure}

As already indicated above, the results obtained so far within the SLN and the approximate approaches LME and QJ are obtained from a factorizing initial state
\begin{equation}\label{factorize}
 W(0)=\rho({0}) \otimes\frac{ {\rm e}^{-\beta H_{R}}}{Z_R}
\end{equation}
 with a quasi-thermalized state $\rho(0) = {\rm e}^{-\beta H_{S}(0)}/Z_{0}$ in (\ref{eq:ini}). This is not the true correlated thermal equilibrium state which the dynamics approaches asymptotically in the absence of driving.  As we have already seen in Fig.~\ref{fig:hf1} and as has also been discussed recently \cite{Ank2014}, the full dynamics thus induces an initial heat flux, even without driving, to establish proper system-bath correlations. Apparently, this happens at a first stage on  relatively short time scales. Accordingly, as seen in Fig.~\ref{fig:therm}, the fully thermalized expectation value $\EW{\sigma_x}$ differs substantially from the quasi-thermalized one $|\langle\sigma_x\rangle_{\beta, 0}|=(\hbar\Delta/2) {\rm tanh}(\Delta\hbar\beta/2)$ even for rather weak coupling.
%
%
Now, in an actual experiment the state of the compound is typically a correlated thermal equilibrium $\rho_\beta={\rm Tr}\{{\rm e}^{-\beta H(0)}/Z\}$. The preparation according to the TMP then provides initial densities projected onto system eigenstates $|k\rangle$ of the form $\rho(0)=\sum_k {\rm Tr}_R\{|k\rangle\langle k|\, {\rm e}^{-\beta H(0)}/Z\}$. The initial state specified in (\ref{eq:ini}) describes this state only in case of extremely weak system-bath interaction.
%
\begin{figure}
\includegraphics[width=0.7\linewidth]{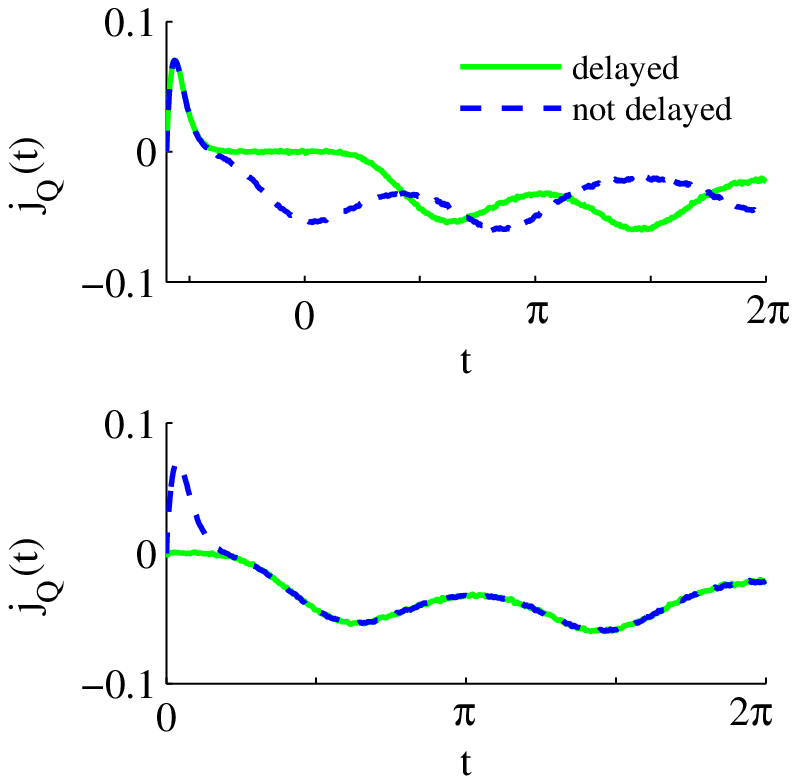}
\includegraphics[width=0.7\linewidth]{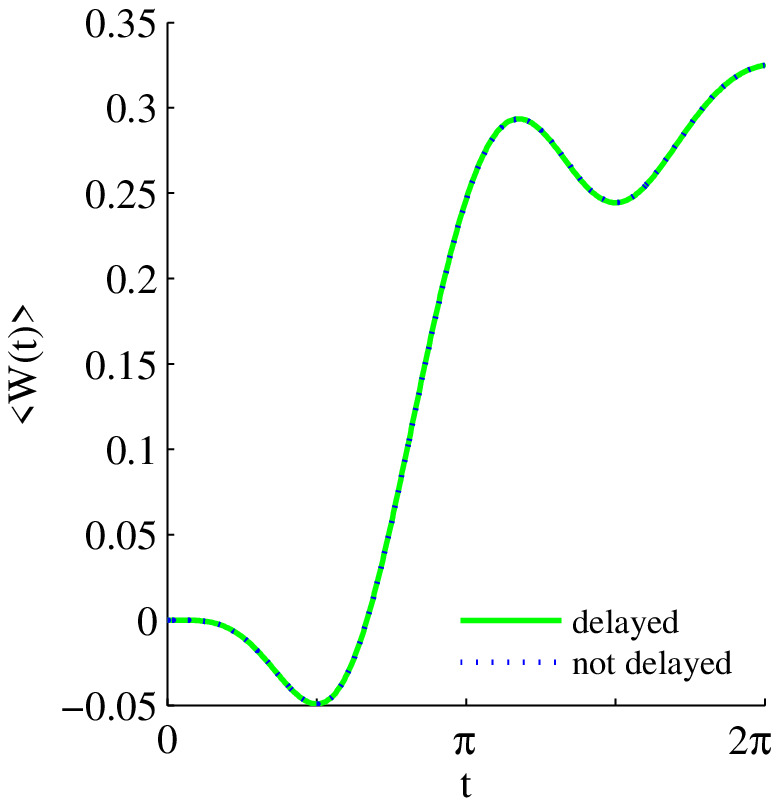}
\caption{\label{fig:comp_delayed} Heat flux (top two panels) and work (bottom)  without (dotted, blue) and with delayed (solid, green) driving for $\beta=1, \lambda_0=1$. Top: Propagation starts from a factorizing initial state with driving acting within $[-0.6\pi, 2\pi]$ (no delay) and within $[0,2\pi]$ (with delay). Middle: Driving within $[0,2\pi]$ starting from a factorizing initial state (dotted, blue) and a correlated initial state (solid, green). Bottom: Work for factorizing initial state with the initial heat pulse subtracted (dotted, blue) and for a correlated initial state (solid, green).}
\end{figure}
While the SLN as it has been formulated in (\ref{sln}) assumes a factorizing state initially, it describes the full equilibration process though. This is not the case for the approximate formulations which assume $\mathcal{W}(t)=\rho(t) \otimes {\rm e}^{-\beta H_R}/Z_R$ for all times. This causes several questions, for example: How reliable are these approximate methods to predict the work from monitoring the change in internal energy and the exchanged heat? As we have seen above, the heat associated with initial correlations is of the same order of magnitude as the heat exchange due to driving meaning that assuming factorized initial states may completely spoil theoretical predictions. Another question then is: Is it possible to separate the time scale on which correlations are established from those on which driving related phenomena occur? According to Figs.~\ref{fig:comp_delayed}, \ref{fig:delayed}, it seems that at least for weak to moderate driving this separation really exists so that one may write $Q=Q_{\rm corr}+Q_D$, where $Q_{\rm corr}$ is the heat due to missing initial correlations and $Q_D$ heat due to driving. Predictions for work and heat based on approximate methods may thus be quantitatively correct for weak driving and weak system-bath coupling if a constant bias, unknown in these approaches though, is subtracted. However, as one observes in Fig.~\ref{fig:delayed}, this no longer applies in the deep quantum regime and for stronger driving, where the relevant time scales tend to overlap. The same is true for stronger system-reservoir coupling. In these situations, a non-perturbative method such as the SLN is mandatory: One first evolves the system for some time in absence of driving starting from a factorizing state until equilibration sets in. Then, this correlated state is properly projected onto system eigenstates and its dynamics in presence of the drive is monitored. Further details will be discussed elsewhere.
%
%

\section{Discussion}\label{sec6}
Based on an exact approach to simulate non-Markovian open quantum dynamics, we study a two level system far from thermal equilibrium in the context of work and heat production. In order to do so, the equivalence of the TMP and the formulation of work in terms of the power operator with properly defined initial states for the first two moment of work is exploited. These results are compared to those obtained with perturbative methods which rely on weak system-bath coupling and Markovian dynamics, and are restricted by the rotating wave approximation. While these latter formulations may at least qualitatively reproduce exact results, sometimes even beyond the strict limits of their applicability, they do fail for other sets of parameters where non-Markovian dynamics and driving are strongly correlated. Therefore, an exact treatment of the open system dynamics is necessary, which we provide here. This approach also allows to retrieve the heat flux and the corresponding heat while the system is driven. It turns out the heat exchange between system and its environment substantially depends on the initial state, at least on a transient time scale. This makes the comparison with experimental data challenging when the associated heat is not known.

While in quantum optical set-ups, one typically works in domains where the system-bath couplings are very weak, this is not always the case for solid state circuits. Further, in the context of superconducting devices strong driving with external microwaves has become an interesting field on its own \cite{wilson:2011,hakonen:2012,stace:2013,gramich:2014} . For future experimental realizations in this direction, predictions of non-perturbative phenomena may thus become very important. While work can be obtained according to the TMP, an alternative procedure is to monitor the change in internal energy and the heat flux. Current experimental developments using ultra-sensitive thermometry in the MHz range follow exactly this strategy \cite{cryo2,cryo3}. This may open a new field to analyze system-bath correlations and even implement well-controlled heat engines. The theoretical framework we applied here, provides the starting point to capture also these settings.

\acknowledgements{
Funding by the DFG through AN336/6-1 (J.A. and R.S.), DAAD/Mincyt (J.A., M.F.C., R.S.), the Foundational Questions Institute (Grant FQXi-RFP3-1317)
(R.S.), and the 7th European Community Framework Programme under grant agreement no. 308850 (INFERNOS), the Academy of Finland (projects 139172 (J.P.), 272218 (J.P.) and 251748 (S.S.)) and the V\"ais\"al\"a Foundation (S.S.) is gratefully acknowledged. We further thank the Universidad Nacional de General Sarmiento, Buenos Aires, (J.A. and R.S.) and the Low Temperature Laboratory at Aalto University, Helsinki, (J.A.) for their kind hospitality. B. Kubala and J. Stockburger are acknowledged for fruitful discussions.}\\


\bibliographystyle{apsrev4-1}
\bibliography{Lit_Widoqs}%
\end{document}